\documentclass[pra,10pt,showpacs]{revtex4}
\usepackage{graphicx,amsmath,amssymb}

\newcommand{\beq}{\begin{equation}}
\newcommand{\eeq}{\end{equation}}
\newcommand{\ket}[1]{| #1 \rangle}
\newcommand{\bra}[1]{\langle #1 |}

\newcommand{\expect}[1]{\langle #1 \rangle}

\newcommand{\ob}[1]{\hat{#1}_i}
\newcommand{\var}{\delta^2}
\newcommand{\UA}{U_A}
\newcommand{\UB}{U_B}
\newcommand{\cov}{C}
\newcommand{\sig}[1]{\hat \sigma_#1}
\newcommand{\LL}{L_2}
\newcommand{\LLL}{L_3}
\newcommand{\MLLL}{{ML}_3}
\newcommand{\MLL}{{ML}_2}
\newcommand{\singlet}{\psi^-}
\newcommand{\noise}{\hat \chi}
\newcommand{\hrk}{\hat {\rho}_k}
\newcommand{\hr}{\hat \rho}
\newcommand{\bellstate}{{\rm Bell}}

\newcommand{\bella}{\psi^+}
\newcommand{\bellb}{\psi^-}
\newcommand{\bellc}{\phi^+}
\newcommand{\belld}{\phi^-}
\newcommand{\bellapm}{\psi^\pm}
\newcommand{\bellspm}{\phi^\pm}

\newcommand{\varr}{\delta^2_{\hat \rho}}
\newcommand{\vark}{\delta^2_{\hat {\rho}_k}}
\newcommand{\abs}[1]{\left| #1 \right|}
\newcommand{\covr}{C_{\hat \rho}}

\newcommand{\brac}[1]{\{ #1 \}}
\newcommand{\arb}{\hat S}
\newcommand{\pmr}{\pm_{\hat \rho}}
\newcommand{\id}{\left( \begin{array}{cc} 1 & 0 \\ 0 & 1 \end{array} \right) }

\begin{document}

\title{Local uncertainty relations serving as measures of entanglement in a
bipartite two-level system}

\author{Simon Samuelsson}
\affiliation{School of Information and Communication Technology,
Royal Institute of Technology (KTH), Electrum 229, SE-164 40 Kista,
Sweden}

\author{Gunnar Bj\"{o}rk}
\affiliation{School of Information and Communication Technology,
Royal Institute of Technology (KTH), Electrum 229, SE-164 40 Kista,
Sweden}

\date{\today}

\begin{abstract}
We comment on the recent suggestion to use a family of local
uncertainty relations as a standard way of quantifying entanglement
in two-qubit systems. Some statements made on the applicability of
the proposed ``measures'' are overly optimistic. We exemplify how
these specific ``measures'' fall short, and present a minor
modification of the general theory which uses the same
experimentally gathered information, but in a slightly different,
better way.

\end{abstract}


\maketitle

\section{Introduction}
During the last few years an interesting approach of using sums of
variances of local observables to probe for the existence of
entanglement in N-level systems has been suggested. The original
proposal by Hofmann and Takeuchi \cite{HT} has appealing features.
Especially, the experimental effort needed to verify that a given
source produces entangled states could be reduced substantially
compared to the full state tomography
\cite{Vogel,Leonhardt,Raymer,Wineland,White,Lvovsky,Barbieri}, joint
measurements of all members of an ensemble
\cite{Peres,Massar,Derka}, or adaptive measurements
\cite{Brody,Fisher}. An extension of the theory has been made by
G\"{u}hne \cite{G}. In addition, Khan and Howell \cite{KH} have
recently investigated the usefulness of the method for bipartite
two-level systems, e.g., consisting of pairs of photons from a
down-conversion source, entangled in polarization. Using
inequalities based on sum-variances measured in different bases on
the singlet state and on mixtures of the singlet state and two kinds
of ``noise'', they state: ``We suggest that these sum-variance
inequalities could be very useful as standard entanglement measures
for spin-1/2 systems.''

In this brief expos\'{e}, we revisit the local uncertainty
relations, or LURs, showing how bipartite entanglement \cite{P} can
be revealed through local-observable correlations. Next, the
proposed ``measures'' of entanglement are reiterated and commented,
and it is shown that they are not suitable to quantify entanglement
outside the frame set by Khan's and Howell's specific investigation.
In this respect they are akin to entanglement witnesses
\cite{Horodecki,Lewenstein,Lewenstein 2,Terhal,Bruss,Bourennane} in
that a LUR can detect certain classes of entangled states, but most
entangled states go undetected. (That is, violation of a LUR is a
sure sign of entanglement, but nonviolation of a LUR implies
nothing.) The general theory of LURs is then modified through the
introduction of modified local uncertainty relations (MLURs) with
improved characteristics.

\section{Local uncertainty relations}
In the initial formulation by Hofmann and Takeuchi [1], LURs
involving two systems, A and B, are considered. In order to
construct a LUR, two sets of observables must be chosen,
$\{\ob{A}\}$ and $\{\ob{B}\}$, acting solely on system A and B,
respectively. Varying the state of the composite system, the sums of
the local variances $\sum_{i} \var \ob{A}$ and $\sum_{i} \var
\ob{B}$, where, e.g., $\delta^2 \ob{A} \equiv \langle \ob{A}^2
\rangle - \langle \ob{A} \rangle^2$, will each have a greatest lower
bound, $\UA$ and $\UB$ respectively. The local uncertainty relation
\beq \sum_i \var{(\ob{A} + \ob{B})} \geq \UA + \UB \label{LUR}, \eeq
is a relation that is supposed to hold for all mixtures of product
states $\hr = \sum_k p_k \hrk$, where $\sum _k p_k = 1$ and $\hrk$
denote product states. The relation is obvious for product states,
since in the expansion of the variance of a sum \beq \var{(\ob{A} +
\ob{B})} = \var{\ob{A}} + \var{\ob{B}} + 2\cov(\ob{A},\ob{B}),
\label{varsum} \eeq the covariance term,
$\cov(\ob{A},\ob{B})=\langle \ob{A} \ob{B} \rangle - \langle \ob{A}
\rangle \langle \ob{B} \rangle$, is simply zero. The extension of
(\ref{LUR}) to hold for a statistical mixture of product states is
easily justified through the general fact that if $\hr = \sum_k p_k
\hrk$, where $\hrk$ represents a product state $\forall \ k$, then
\beq \varr \arb \geq \sum_k p_k \vark \arb \label{mixvar} \eeq for
an arbitrary observable $\arb$, adding a subscript to the variances
to make their state dependence explicit. The inequality
(\ref{mixvar}) holds because $\varr \arb = \sum_k p_k \expect{(\arb
- \expect{\arb}_{\hr})^2}_{\hrk} = \sum_k p_k \brac{\vark \arb +
(\expect{\arb}_{\hrk} - \expect{\arb}_{\hr})^2} \geq \sum_k p_k
\vark \arb$.

When the systems are mixed or entangled, the covariance term in
(\ref{varsum}) is in general nonzero. If the LUR is to reveal
nonseparability, a necessary condition is that at least one of the
covariances between the systems are such that \beq
\cov(\ob{A},\ob{B}) < 0 . \label{sumcov} \eeq The covariance is
bounded from below and above by the following relation: \beq
-(\var\ob{A} + \var\ob{B}) \leq 2\cov(\ob{A},\ob{B}) \leq \var\ob{A}
+ \var\ob{B} \label{covbound}. \eeq Interestingly, for any
particular choice of a pair of observables $\ob{A}$ and $\ob{B}$,
both bounds can be reached both with mixed separable, and pure
entangled states. Hence, the essential difference between separable
and entangled states cannot be distinguished by studying one
observable pair alone, but lies rather in the existence of
significant covariance between several pairs of observables
simultaneously. The system of two spin-1/2 particles serves as an
illustrative example. Measuring spin in the three cartesian
directions $x$, $y$ and $z$, the revelation of maximum covariance in
one direction implies, for separable states, no covariance in the
other two directions. On the other hand, entangled states may
possess maximum covariance [it saturates one of the bounds in
(\ref{covbound})] in all three bases. The LURs exploit this
characteristic feature of entanglement. However the important
observation that follows from condition (\ref{sumcov}) is that {\em
a given LUR can only detect the entanglement for states with
``appropriate'' signs of the covariances}.

Focussing on two-level systems, through polarization states of
photon pairs, Khan and Howell investigated two closely related
LURs based on polarization measurements in different bases: \beq
\LL = \var{(\sig{A} + \sig{B})_{0/90}} + \var{(\sig{A} +
\sig{B})_{45/135}} \geq 2 \label{LL} \eeq and \beq \LLL =
\var{(\sig{A} + \sig{B})_{0/90}} + \var{(\sig{A} +
\sig{B})_{45/135}} + \var{(\sig{A} + \sig{B})_{R/L}} \geq 4
\label{LLL} .\eeq The lower bounds $\LL \geq 2$ and $\LLL \geq 4$
are valid for any mixture of product states but may be violated
for entangled states. The subscript 0/90 refers to a measurement
of horizontal-vertical polarization, while for 45/135, the basis
is rotated by 45 degrees. The third term to the left of the
inequality sign of (\ref{LLL}) denotes right- and left-handed
circular polarization. In these two cases, $\ob{A}$ and $\ob{B}$
correspond to identical measurements on the two systems, so in the
light of condition (\ref{sumcov}), negative covariances are
favored. The optimum domain of entanglement verification for these
LURs therefore necessarily involves the singlet state,
exemplified, e.g., by a superposition of vertically (V) and
horizontally (H) linearly polarized photons: $\ket{\singlet} =
(\ket{H,V} - \ket{V,H})/\sqrt{2}$. This state has the unique
property among the maximally entangled states that a measurement
of $\ob{A} + \ob{B}$, where $\ob{A}$ and $\ob{B}$ represent
arbitrary, but identical polarization or spin measurements on the
two particles, will always give the trivial outcome zero with
certainty.

The only states previously studied \cite{HT,G,KH} have density
operators of the form \beq \hr = p\ket{\singlet}\bra{\singlet} +
(1-p)\noise \label{states}, \eeq where $0 \leq p \leq 1$. These are
clearly based on the singlet state, and the density operator
$\noise$ refers to either Werner noise or maximally polarized noise
\cite{P}. For these states both $\LL$ and $\LLL$, the latter in
particular, shows high sensitivity in detecting entanglement,
compared to the corresponding Bell measurement. Analytically it has
been shown that entanglement exists for values of $p$ within the
range $1/3 < p \leq 1$ for Werner noise and $0 < p \leq 1$ for
maximally polarized noise. $\LLL$ drops below its lower bound 4
precisely for these values of $p$.

The application of these particular LURs has been extended further
by letting the magnitude of the violation constitute a measure of
the degree of entanglement for a given state. Khan and Howell
\cite{KH} even suggest application beyond the states of the form
(\ref{states}) concerning $\LLL$ by claiming: ``This measurement
therefore should be ideal for standardizing entanglement measures
in spin-1/2 systems''. This extended use, however, requires that
the behavior of the LURs obey certain criteria. One important such
is invariance under local unitary transformations. However, such
transformations are capable of changing both the amount and the
sign of the covariance (due to entanglement) in different bases,
while they are unable to alter the degree of entanglement in the
system. As shown by Eq. (\ref{varsum}), a single LUR is highly
sensitive to the kind of covariances present. This unfortunately
makes them unsuitable as entanglement measures for general states.

As an example, the four Bell states $$ \ket{\bellapm} =
\frac{1}{\sqrt{2}} \left (\ket{H,V} \pm (\ket{V,H}\right) \quad
{\rm and}$$ $$ \ket{\bellspm} = \frac{1}{\sqrt{2}} \left
(\ket{H,H} \pm (\ket{V,V}\right) $$ can be considered. Since each
Bell state can be transformed by a local unitary transformation
into any of the other, a functional associating each state in the
Hilbert space with a real value can only be considered a
reasonable measure of entanglement if all the Bell states are
assigned the same values. This is not the case for the
``measures'' $\LL$ and $\LLL$. Only the singlet state gives the
desired values $\LL = \LLL = 0$ of maximum violation, and for the
other three Bell states, $\LL = 4$ and $\LLL = 8$, which is not
even close to violating the corresponding LURs, in spite of the
fact that these states are maximally entangled. The reason for
this is that these Bell states have negative polarization
covariance in one basis only, and positive covariance in the other
two. Consequently, $\LL$ and $\LLL$ will only detect entangled
states belonging to the class defined by Eq. (\ref{states}).

\section{Modified local uncertainty relations}
Fortunately, with a slight modification the LURs can be restated in
a manner that makes them sensitive only to the magnitude, and not to
the sign, of the covariances, requiring no further experimental
effort. For ordinary LURs, the expression $\sum_i \var{\ob{A}} +
\var{\ob{B}} + 2\cov(\ob{A},\ob{B})$ is bounded below, and derives
from expanding (\ref{LUR}) using (\ref{varsum}). Now forcing all the
covariance terms to contribute negatively by subtracting the
correlation moduli instead, we arrive at the modified LUR, or MLUR,
\beq \sum_i \var \ob{A} + \var \ob{B} - 2
\abs{{\cov(\ob{A},\ob{B})}} = \sum_i \var{(\ob{A} + \ob{B})} - 4\
{\rm Max}\brac{0,\cov(\ob{A},\ob{B})} \geq \UA + \UB. \label{ML}
\eeq In any measurement where variances of the form $\var (\ob{A} +
\ob{B})$ are determined, the rates of incidence necessary to
calculate $\var \ob{A}$ and $\var \ob{B}$ are also directly
available. Therefore, the covariances in (\ref{ML}) are at hand
through Eq. (\ref{varsum}).
    A vital question is whether the lower bound for separable states,
$\UA + \UB$, is still valid for the MLURs, as already implied. To
prove that this is indeed the case we consider the first sum in
(\ref{ML}) for a mixture of product states $\hr$. Since the
covariance is linear in both arguments we can choose a sign $\pmr$
preceding $\ob B$, depending on the state $\hat \rho$, so that
each term in (\ref{ML}) can be written as \beq \varr \ob{A} +
\varr (\pmr \ob{B}) + 2 {\covr(\ob{A},\pmr \ob{B})} = \varr(\ob{A}
\pmr \ob{B}) \geq \sum_{k} {p_k \vark(\ob {A} \pmr \ob{B})} =
\sum_{k} {p_k \vark(\ob {A} + \ob{B})}. \eeq The inequality is
nothing but (\ref{mixvar}), and the last equality holds only
because $\hrk$ denotes a product state, and the covariance term is
zero for product states. Thus, the inequality (\ref{ML}) becomes

\beq \sum_i \varr \ob{A} + \varr \ob{B} - 2
\abs{{\covr(\ob{A},\ob{B})}} \geq \sum_i \sum_{k} {p_k \vark(\ob {A}
+ \ob{B})} \geq \sum_k p_k(\UA + \UB) = \UA + \UB. \eeq We conclude
that the modification of a LUR into an MLUR does not alter the
greatest lower bound for separable states.

\begin{figure}
\centering
\includegraphics[width=8cm]{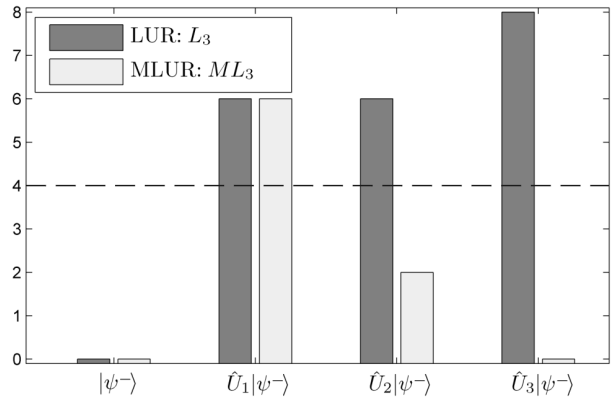}
\caption{The difference in behavior of a LUR and an MLUR,
exemplified through $\LLL$ and $\MLLL$, for fully entangled
states.}
\end{figure}

This modification of the theory inherits all the appealing
characteristics of the LURs, while extending the subset of
entangled states that can be detected with any given LUR, as seen
in Fig. 1. The ideal case is maximum violation of the lower bound
4, fulfilled by both expressions for the singlet state
$\ket{\singlet}$. However, for the local unitary transformation
$\hat {U}_1 \ket{\singlet} = (i\ket{\bella} + \ket{\bellb} +
\ket{\bellc} -i\ket{\belld})/2$, $\MLLL$ experiences its
worst-case scenario under all such transformations, where all
three covariances vanish and $\MLLL = \LLL = 6$. On the other
hand, the improved characteristics of MLURs over LURs are
illustrated by the states $\hat {U}_2 \ket{\singlet} = (-
i\sqrt{3} \ket{\bella} + \ket{\bellb})/2$ and $\hat {U}_3
\ket{\singlet} = \ket{\bella}$ for which $\LLL$ is unable to
detect entanglement. $\MLLL$ verifies the existing maximum
entanglement through partial and full violation, respectively, of
its bound. The explicit form of the local unitary transformations
used above are $$\hat {U}_1 = \frac {1}{2} \left(
\begin{array}{cc} 1+i & -1+i \\ 1+i & 1-i
\end{array} \right) \otimes \id, $$ $$ \hat {U}_2 = \frac
{1}{2} \left( \begin{array}{cc} 1 - i\sqrt{3} & 0 \\ 0 &
1+i\sqrt{3}
\end{array} \right) \otimes \id,  \quad {\rm and} $$  $$ \hat {U}_3 = \left(
\begin{array}{cc} 1 & 0 \\ 0 & -1 \end{array} \right) \otimes \id
. $$

The modification of $\LLL$ into $\MLLL$ amounts to treating all
Bell states in the same manner, thus avoiding the troublesome
behavior mentioned earlier. This new MLUR is in fact maximally
violated for all four Bell states. As a consequence, this
generalizes the results regarding entanglement verification for
states of the form (\ref{states}). The MLUR is similarly violated
for the kinds of noise studied, if the singlet states are replaced
by any other Bell state, that is $\hr =
p\ket{\bellstate}\bra{\bellstate} + (1-p)\noise$.

The concept of MLURs also provides useful insight in the way $\LL$ 
detects entanglement. Given any state, $\LL$ is always non-negative and 
bounded above by the inequalities
$0 \leq \LL \leq 8$, and the existence of entanglement is ensured for  
$0 \leq \LL < 2$. However, significant covariances in the two particular bases
used must also be present in order to find $\LL$ within the range 
$6 < \LL \leq 8$. By constructing $\MLL$, the MLUR corresponding to $\LL$, all 
states satisfying this condition on $\LL$ are found to violate the MLUR. 
 That is, $\MLL < 2$ if $\LL > 6$. In order to prove this statement, we first note that 
$\var{\ob{A}} \leq 1$ for any polarization measurement performed on one of the
photons. The explicit expressions for $\MLL$ and $\LL$ each contain four such 
variances. Thus, using the definitions of $\LL$ and $\MLL$ through the 
equality in (\ref{LL}) and the left-hand sum in (\ref{ML}), respectively, 
$$\MLL \leq 4 - \abs{2\cov{(\sig{A},\sig{B})_{0/90}} + 2\cov{{(\sig{A},\sig{B})_{45/135}}}}
\leq 4 - (\LL - 4) < 2, $$ keeping in mind that $\LL > 6$. 
This fact consequently introduces an upper bound in the LUR in Eq. (\ref{LL}), so that
$2 \leq \LL \leq 6$ for separable states. This would be of immediate use
for anyone clinging to $\LL$. However, we advocate the use
of $\MLL$ instead of $\LL$. The reason for this is simply that the MLUR 
can detect entanglement for an even larger class of states, namely those having
significant covariances with opposite signs. For such states $\LL$ typically
fails to detect entanglement. The state $\hat {U}_2 \ket{\singlet} = (-i\sqrt{3} \ket{\bella} + \ket{\bellb})/2$
exemplifies this, having $2\cov{(\sig{A},\sig{B})_{0/90}} = -2$
and $2\cov{{(\sig{A},\sig{B})_{45/135}}} = 1$, since $\LL = 3$, providing no information, 
but $\MLL = 1$, verifying present entanglement.

It must be stressed, however, that not even the suggested MLURs should be
considered as measures of entanglement for general states. There are
still certain local unitary transformations of the Bell states,
e.g. $\hat {U}_1 \ket{\singlet}$ in Fig. 1, such that the
transformed states do not reveal their entanglement through the
particular observables used in e.g. $\LLL$ and $\MLLL$. Rather, LURs
and MLURs should be used with the same caution as we do with
entanglement witnesses. MLURs cannot detect all kinds of bipartite
entanglement, but on the other hand, the necessary measurements
can be performed with much less resources than, e.g., a full state
tomography (that, on the contrary, will detect entanglement,
irrespective of its type).

\section{Conclusions}
Conclusively, using the notion of covariance, we have shown how
entangled states can violate LURs. However it also reveals the
dependence on the sign of the covariance, causing an unnatural
asymmetry in the way Bell states are treated by the proposed
measures $\LL$ and $\LLL$ \cite{KH}. Using this information, the
general theory of LURs has been refined by introducing MLURs that
process identically the same measurement data in a more efficient
way. Their advantage lies in the fact that they can detect
entanglement for a larger class of states than the corresponding
LURs, without any additional measurements. In the specific case of
$\LLL$ and $\MLLL$, their relative ease to measure is at the
expense of both being sensitive to local unitary transformations.
Computer simulations show better behavior of $\MLLL$ than $\LLL$,
although it can be deceptive to regard any of them as a general
measure of entanglement. Instead, $\MLLL$ should be viewed as a
measure of ``Bell-type'' entanglement, which is of substantial
interest in characterizing the quality of entanglement from
different sources.

\acknowledgments{This work was partially supported by the Swedish
Foundation for Strategic Research (SSF) and the Swedish Research
Council (VR).}

\end{document}